\documentclass[prl,floatfix,twocolumn,aps,amsmath,amssymb,nofootinbib,preprintnumbers]{revtex4}
\usepackage{graphicx}
\usepackage{bm}         % bold math
\usepackage{amsmath}
\usepackage{amsfonts}
\usepackage{color}

\definecolor{dg}{rgb}{0.,0.5,0.}

\begin{document}

\title{Collisional Penrose process near the horizon of extreme Kerr black holes}

\author{Micha{\l} Bejger$^1$\footnote{E-mail address: {\tt bejger@camk.edu.pl}},
Tsvi Piran$^2$\footnote{E-mail address: {\tt tsvi.piran@mail.huji.ac.il}},
Marek Abramowicz$^{1,3}$\footnote{E-mail address: {\tt marek.abramowicz@physics.gu.se}},
Frida H{\aa}kanson$^4$}

\affiliation{
$^1$Copernicus Astronomical Center, ul. Bartycka 18, 00-716, Warszawa, Poland\\
$^2$Racah Institute of Physics, The Hebrew University, Jerusalem 91904, Israel\\
$^3$Department of Physics, University of Gothenburg, 412-96, G{\"o}teborg, Sweden\\
$^4$Department of Physics, Chalmers Technical University, 412-96, G{\"o}teborg, Sweden
}
\date{\today}

\begin{abstract}
Collisions of particles in black holes' ergospheres may result in an arbitrarily large
center-of-mass energy. This led recently to the suggestion (Ba\~nados et al., 2009) that
black holes can act as ultimate particle accelerators. If the energy of an outgoing particle
is larger than the total energy of the infalling particles, the energy excess must come from
the rotational energy of the black hole and hence a Penrose process is involved. 
However, while the center-of-mass energy diverges, the position of the collision makes it
impossible for energetic particles to escape to infinity. Following an earlier work on
collisional Penrose processes (Piran \& Shaham 1977), we show that even under the most
favorable idealized conditions the maximal energy of an {\it escaping particle} is only a
modest factor above the total initial energy of the colliding particles. This implies that
one shouldn't expect collisions around a black hole to act as spectacular cosmic
accelerators.
\end{abstract}

\preprint{...}
\maketitle

In a seminal paper, Penrose \citep{Penrose1969} suggested the possibility of extracting the
rotational energy of a black hole (BH). Penrose, and later Penrose and Floyd \cite{PF71}, 
considered an infalling particle disintegration in the ergosphere of a Kerr BH. One 
of the particles produced in this process might be thrown into a negative energy 
(with respect to infinity) orbit, while the other one will have an energy 
$E_{\rm out}$ larger than the energy $E_{\rm in}$ of the infalling one. 
The energy excess arises eventually from the rotational energy
of the BH.  However, shortly afterwards it was shown \citep{Kovetz_Piran_75,Wald_74}
that in order to have $E_{\rm out}>E_{\rm in}$, the disintegration process must convert most of
the rest-mass energy of the infalling particle to kinetic energy. Such a disintegration
mechanism does not exist in nature for stable particles, rendering the original
Penrose process irrelevant for realistic astrophysical situations.

In a collision between two particles,  the center-of-mass (CM) frame the collisional energy can be mostly kinetic, and outgoing 
particles from within the ergosphere might easily have $E_{\rm out}>E_{\rm in}$, 
resulting in a collisional Penrose process.  In fact further studies, \citep{PS1977},
showed that when two particles collide near the horizon, the CM energy 
$E_{\rm cm}$ can be arbitrarily large (see sect. IIG of \citep{PS1977}). This fact was
recently used by \citep{BSW2009} in the context of collisional dark matter. It was shown
that for the extremal (spin parameter $a=J/M=1$, where $J$ and $M$ are BH angular momentum
and mass, respectively) Kerr BH, and a collision at the outer horizon ($r\to r_{\rm h} =
M + \sqrt{M^2 - a^2}$), there are cases (when one of the colliding particles  has a critical
angular momentum) with $E_{\rm cm}(r\to  r_{\rm h})\to \infty$. Unbounded $E_{\rm cm}$ was
subsequently proposed as an energy source for a Planck-scale particle accelerator.

This idea attracted a lot of attention and resulted in numerous papers repeating the
statement that the CM energy of colliding particles may grow limitlessly in
circumstances different than those considered by \citep{BSW2009}. These variations include
different colliding particles (e.g. charged, massless, spinning), different spacetimes (e.g.
naked singularity, string, non-zero cosmological constant), gravity theories different from 
Einstein's GR, and different collision settings (e.g. plunging from ISCO, collisions at the
inner Kerr horizon). There were also several attempts to provide a ``simple explanation''
for the infinite CM energy.

It seems that the main result of \citep{BSW2009} was generally accepted and, when it was
criticized, the criticism related to rather irrelevant issues, e.g. gravitational radiation,
or self-gravity. Instead, here we present a more substantial criticism of the meaning and
physical significance of the result of \citep{BSW2009} based on the statement that, 
while particles may locally reach huge energies, one has to consider the question of whether
they escape to infinity (\citep{PS1977}). This is not trivial: to be energetic, the collision
has to take place extremely close to the BH which, however, impedes the  
particle's escape. In fact, given the geometry of a typical collision, 
one might expect that the most energetic particles will fall into the BH.

This question was touched on before in a few papers which followed \citep{BSW2009}. These papers
attempted to examine the likelihood of the collision products escaping to infinity
\cite{Berti2010,Jacobson2010,Banados2011,Williams2011,Patil2011}. Here, we go further and
examine the {\it maximal energy} of particles which {\it actually escape} to infinity. These
values, rather than the maximal available energy in the CM frame, should be the relevant
ones for astrophysical considerations.

Specifically, we calculate the upper limit on the energy of an escaping photon which results
from a collision between two infalling particles. The parameter phase space is very large: 
it involves the energy, angular momentum and Carter's constant of the two infalling
particles, the coordinates of the collision point, and the masses and directions of the
outgoing particles. Rather then exploring the whole phase space, we examine the special
case of two particles falling from rest at infinity and colliding in the ergosphere of a Kerr BH. 
The collisions take place in the equatorial plane of the Kerr BH and result 
in two photons which also move in the equatorial plane. Symmetry considerations suggest that
collisions which take place in the equatorial plane would result in the most energetic particles
and that photons will escape most easily from the vicinity of the BH, hence the
upper limit that we find here is the true upper limit for the energy of an escaping particle
from a near-BH collision.

%-----------------------------------------------------------------------------
%%% \section{Introduction and model}
%-----------------------------------------------------------------------------

Following \citep{PS1977} we consider the Kerr metric in the standard 
Boyer-Lindquist coordinates. In these coordinates, the Kerr metric tensor $g_{\mu \nu}$
depends neither on time $t$, nor on azimuthal angle $\phi$. This implies two Killing
symmetries, ${\cal L}_{\eta} g_{\mu \nu} = 0 = {\cal L}_{\xi} g_{\mu \nu}$ given in terms of
the two Killing vectors, which in the Boyer-Lindquist coordinates are $\eta^{\nu} =
\delta^{\nu}_{~t}$ and $\xi^{\nu} = \delta^{\nu}_{~\phi}$. The equations of geodesic motion
admit two non-trivial constants of motion, which may be expressed by the particle's (or photon's)
four-momentum $p^{\nu}$ and the two Killing vectors: energy at infinity $E = -
p_{\nu}\eta^{\nu}$, and angular momentum parallel to the BH axis $L = p_{\nu}\xi^{\nu}$. The
third constant of motion, the Carter constant $Q$, vanishes for particles and photons
moving in the equatorial plane $\theta = \pi/2$.

We will make use of two local frames of reference to study the physical properties of the
collision. The first one is the locally non-rotating frame (LNRF, also called ZAMO,
\citep{BardeenPT1972}), and the second one is the center-of-mass frame (CM). Each of these
frames defines its own ``comoving observer'', who is at rest in his corresponding frame.

The family of LNRF observers has its trajectories orthogonal to the space-like hypersurfaces
$t=\,const.$, and their four-velocities given by
%-----------------------------------------------
\begin{equation}
\label{CM-velocity-definition}
N^{\nu} = e^{-\Phi}\left[ \eta^{\nu} + \omega \xi^{\nu} \right],
\end{equation}
%-----------------------------------------------
with $\Phi$$\,=\,$$\ln[-(\eta^{\nu}\eta_{\nu})$$\,-\,$$\omega(\xi^{\nu}\eta_{\nu})]^{1/2}$,
$\omega
$$\,=\,$$-(\xi^{\nu}\eta_{\nu})/(\xi^{\mu}\xi_{\mu})$.

For two particles with the four momenta $p^{\mu}_{~(1)} = m_{(1)} u^{\mu}_{~(1)}$ and
$p^{\mu}_{~(2)} = m_{(2)} u^{\mu}_{~(2)}$, the comoving observer in the CM frame has his
four-velocity $U^{\mu}$ given by the two conditions: the first one is $U^{\mu}\,U_{\mu} =
-1$, and the second one is
%-----------------------------------------------
\begin{equation}
\label{CM-velocity-definition2}
U^{\mu} = \frac{1}{E_{\rm cm}} \left [p^{\mu}_{~(1)} + p^{\mu}_{~(2)} \right ],
\end{equation}
%-----------------------------------------------
where the center-of-mass energy $E_{\rm cm}$  equals
%-----------------------------------------------
\begin{equation}
E_{\rm cm} = \sqrt{m^2_{(1)} + m^2_{(2)} - 2m_{(1)}\,m_{(2)} g_{\mu\nu}u^\mu_{~(1)}
u^\nu_{~(2)}}.
\label{eq:ecm}
\end{equation}
%-----------------------------------------------
Here $m_{(i)}$ are the masses and $u^\mu_{~(i)}$ are the four velocities of the two
($i=1,2$) particles.

\begin{figure}[ht]
\includegraphics[width=\columnwidth]{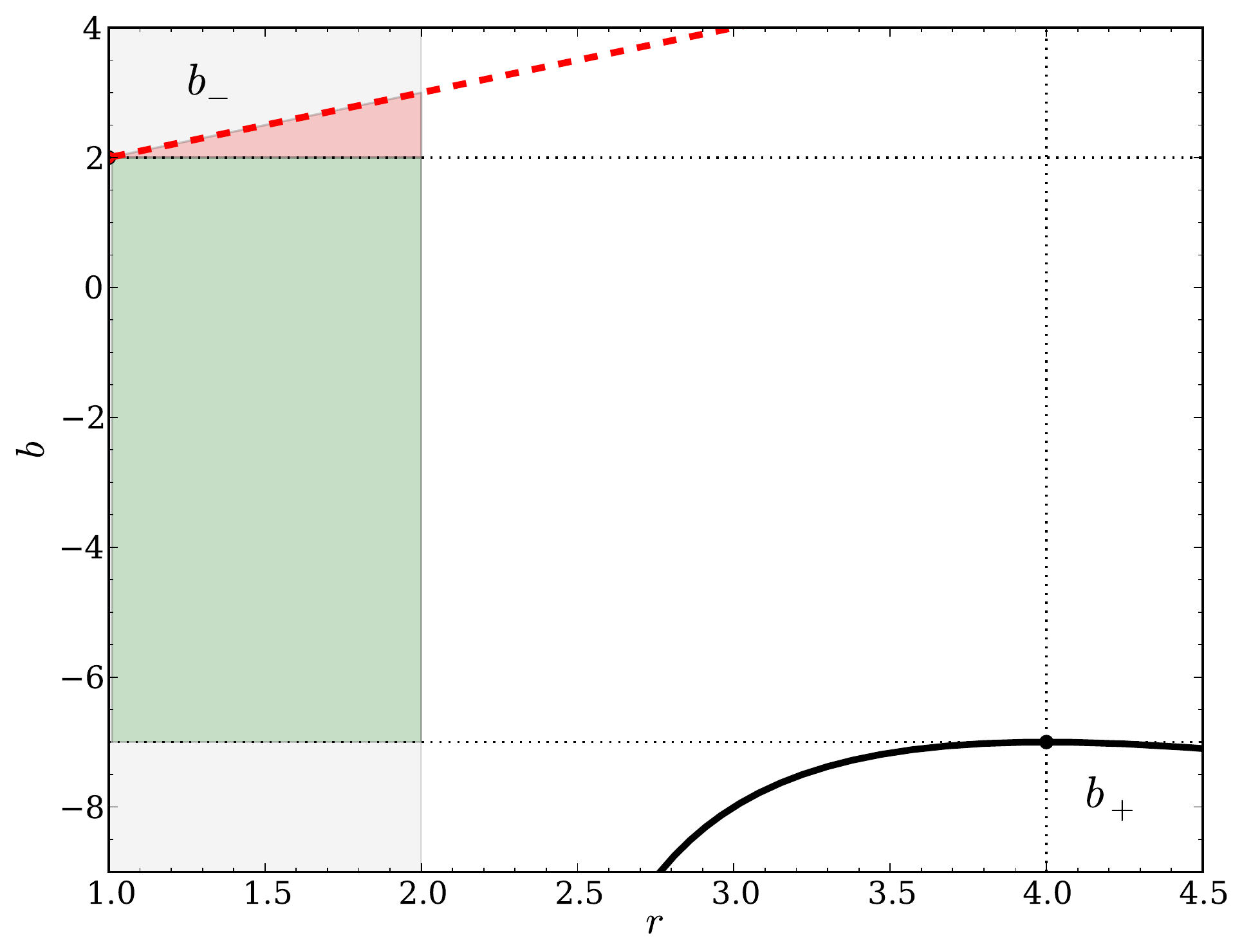}
\caption{Conditions of the photon impact parameter $b=L/E$ in case of $a=1$,
plotted as $b_{+}$ (solid curve) and $b_{-}$ (dashed curve) functions
(Eq.~\ref{eq:bpm}). In order to escape the ergosphere, infalling photons must 
have $b>2$; for initially-outgoing photons, $b>-7$. Grey region denotes 
the ergosphere.}
\label{fig:b_a1}
\end{figure}

The two photons resulting from a collision should have their four momenta given in the CM
frame (\ref{CM-velocity-definition2}) by
%-----------------------------------------------
\begin{eqnarray}
p^{\mu}_{~(3)} &=& \frac{E_{\rm cm}}{2}\left[ U^{\mu} + \tau^{\mu}\right],
\label{photon-plus} \\
p^{\mu}_{~(4)} &=& \frac{E_{\rm cm}}{2}\left[ U^{\mu} - \tau^{\mu}\right],
\label{photon-minus}
\end{eqnarray}
%-----------------------------------------------
where $\tau^{\mu}$ is an (arbitrary) direction in the instantaneous 3-space of the CM observer
(assumed here to be in the equatorial plane, $\theta=\pi/2$). The conserved energies ``at
infinity'' of these photons are,
%-----------------------------------------------
\begin{equation}
\label{photon-energies}
E_{(3)} = -\eta_{\mu}p^{\mu}_{~(3)}, ~~E_{(4)} = -\eta_{\mu}p^{\mu}_{~(4)},
\end{equation}
%-----------------------------------------------
and the respective angular momenta are
%-----------------------------------------------
\begin{equation}
\label{photon-momenta}
L_{(3)} = \xi_{\mu}p^{\mu}_{~(3)}, ~~L_{(4)} = \xi_{\mu}p^{\mu}_{~(4)}.
\end{equation}
%-----------------------------------------------
The subsequent fate of the photon is decided by the point of collision with respect
to the location of the turning points provided by the effective radial potential.
In the Boyer-Lindquist coordinates, it reads
\begin{eqnarray} V_r &=& E^2r^4 + (Era)^2
 - r(r-2M)L^2\nonumber \\
 &-& 4MrELa + 2Mr(Ea)^2.
\end{eqnarray}
For $V_r=0$ one obtains a relation for the photon impact parameter $b=L/E$,
\begin{equation}
b_\pm = \frac{2Ma \pm \sqrt{r^4 -2Mr^3 + (ar)^2}}{2M-r}.
\label{eq:bpm}
\end{equation}
In the case of $a=1$ and for initially infalling photons, the escape
conditions are such that $b>2$. For initially outgoing photons, $b$ must be
greater than the maximum of $b_{+} = -7$, located at $r=4M$
(see Fig.~\ref{fig:b_a1}).

%-----------------------------------------------------------------------------
%-----------------------------------------------------------------------------
%-----------------------------------------------------------------------------
%%%\section{Results}
%-----------------------------------------------------------------------------

We examine here a collision in the ergosphere between two massive ($m_{(1)}=m_{(2)}=1$) test
particles infalling from rest at infinity with $E_{(1)} = E_{(2)} = 1$.  We first consider 
an extremal Kerr BH (spin parameter $a=1$) and  we set the angular momentum
of one of the particles to  $L_{(1)}=2$, the critical $L$ for this value of $a$.  We then vary
the point of collision (distances are measured in units of $M$, the BH's mass), as well as 
the initial angular momentum of the second particle, $L_{(2)}$.

\begin{figure}[ht]
\includegraphics[width=0.5\textwidth]{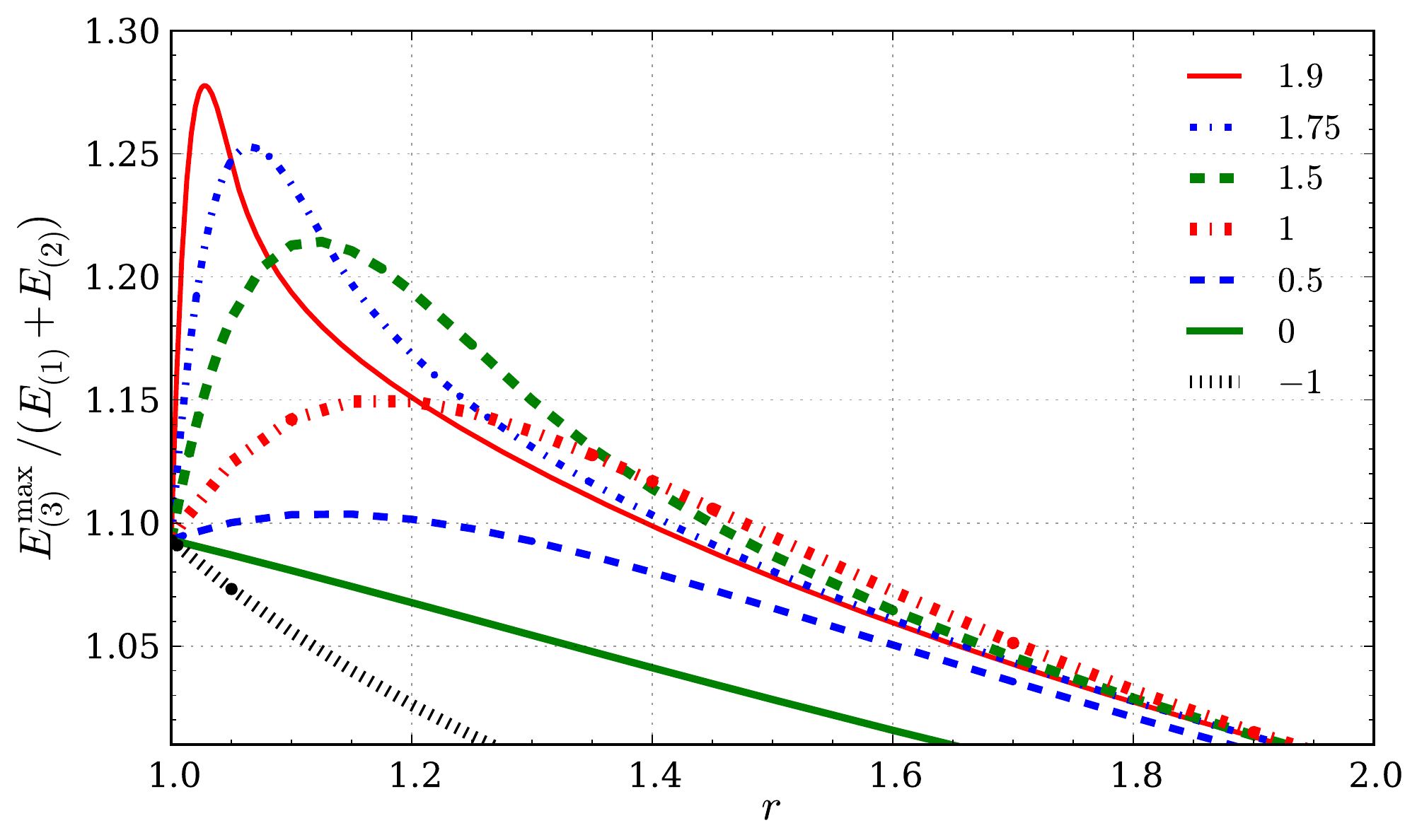}\\
\includegraphics[width=0.495\textwidth]{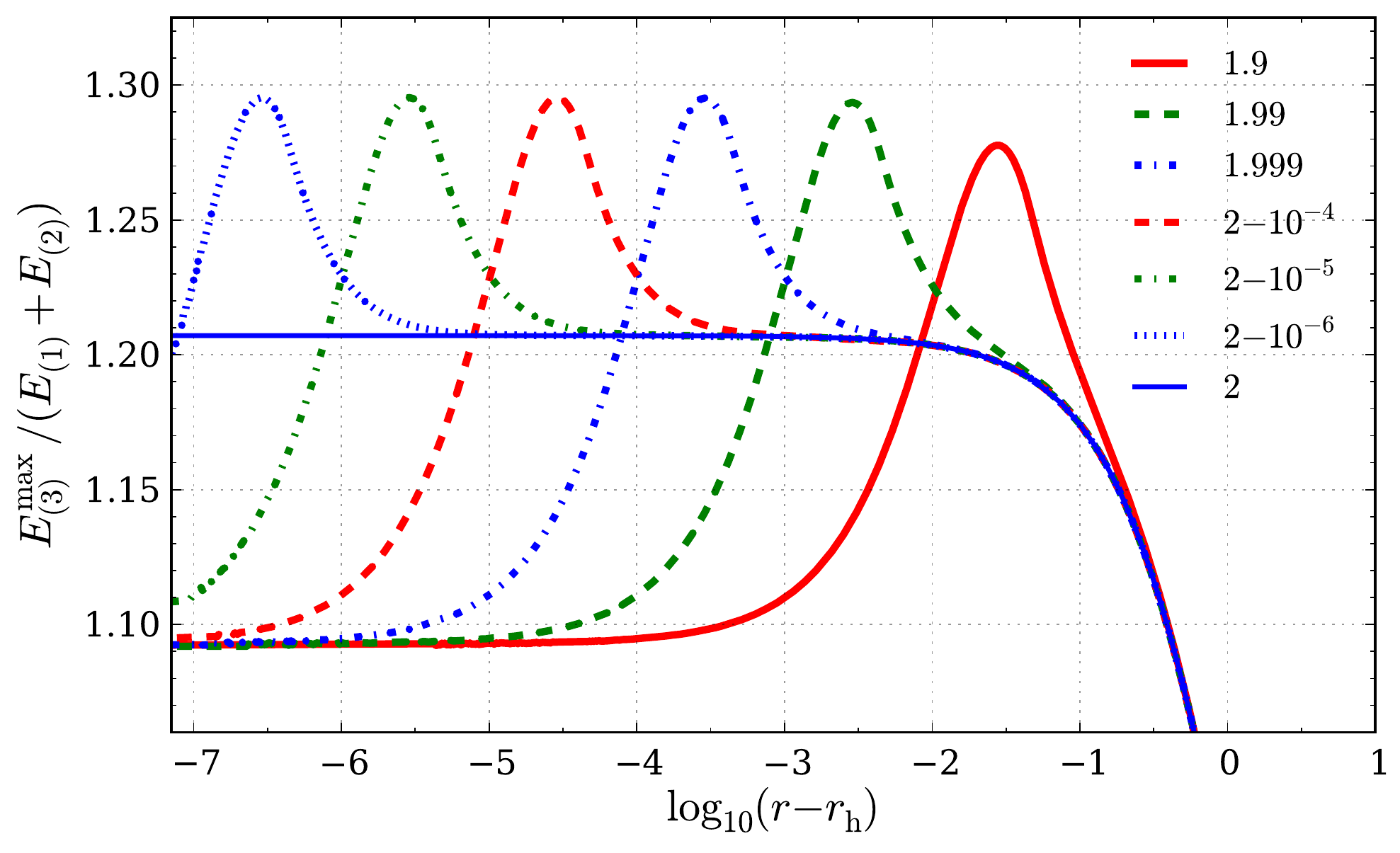}
\caption{{\it Upper panel}: The maximal energy of an escaping, initially
ingoing photon, resulting from a collision of two massive ($m_{(1)}=m_{(2)}=1$)
particles with $L_{(1)}=2$ and $L_{(2)} \in (-1,1.9)$ as a function of the distance, measured in units
of the BH's mass, $M$,
from the BH center ($a=1$). {\it Lower panel}: Zoom for $r\to r_{\rm h}$ and
$L_{(2)}\to 2$.}
\label{fig:a1l12}
\end{figure}

\begin{figure}[ht]
\includegraphics[width=0.5\textwidth]{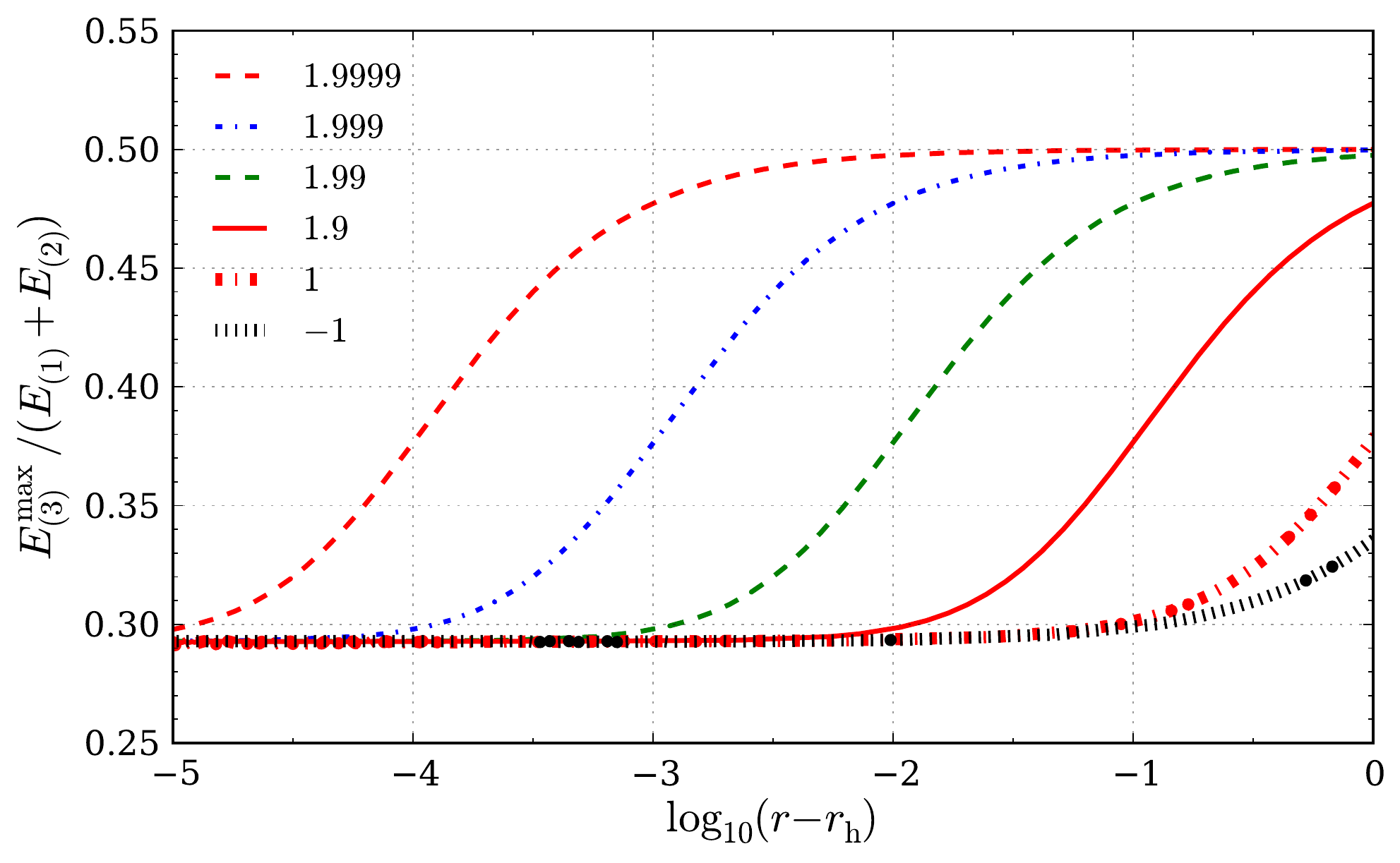}
\caption{Maximal energy of an {\it initially outgoing} photon as a function
of the distance from the BH center, for $L_{(1)}=2$, $L_{(2)} \in (-1,1.9999)$
($a=1$).}
\label{fig:a1l12_mo}
\end{figure}

We find that photons which actually escape the ergosphere with an 
energy gain, $E_{\rm out} > E_{\rm in}$, i.e., $E_{(3)} > E_{(1)} + 
E_{(2)}$, are all {\it initially infalling}, and are deflected just 
before plunging into the BH. Fig.~\ref {fig:a1l12} depicts 
the maximal energy as a function of the point of collision for 
various values of $L_{(2)}$. When the collision approaches the 
horizon, $E^{\rm max}_{(3)}/(E_{(1)} + E_{(2)})$ peaks at $1.295$ at 
a slight distance from the horizon (which approaches the horizon as 
$L_{(2)} \rightarrow 2$) and then declines asymptotically to 1.093 
on the horizon. We stress that in all these 
cases $E_{\rm cm}$ diverges as the collision 
approaches the horizon. The situation is different at the limit  
$L_{(2)} = 2$. Now, the ratio $E^{\rm max}_{(3)}/(E_{(1)} + 
E_{(2)}) \rightarrow (1 + \sqrt{2})/2$ on the horizon. Since both 
particles fall on the same orbit (there is no real collision here) 
the situation coresponds to the case of particle decay \cite
{PS1977,Wald_74}. The maximal energy of {\it initially outgoing} 
photons that escape the ergosphere is presented in Fig.~\ref
{fig:a1l12_mo}. These are not Penrose particles, since $E^{\rm 
max}_{(3)}/(E_{(1)} + E_{(2)})<1$ (close to the horizon, the value 
approaches $1/(2+\sqrt{2})$).

\begin{figure}[ht]
\includegraphics[width=0.5\textwidth]{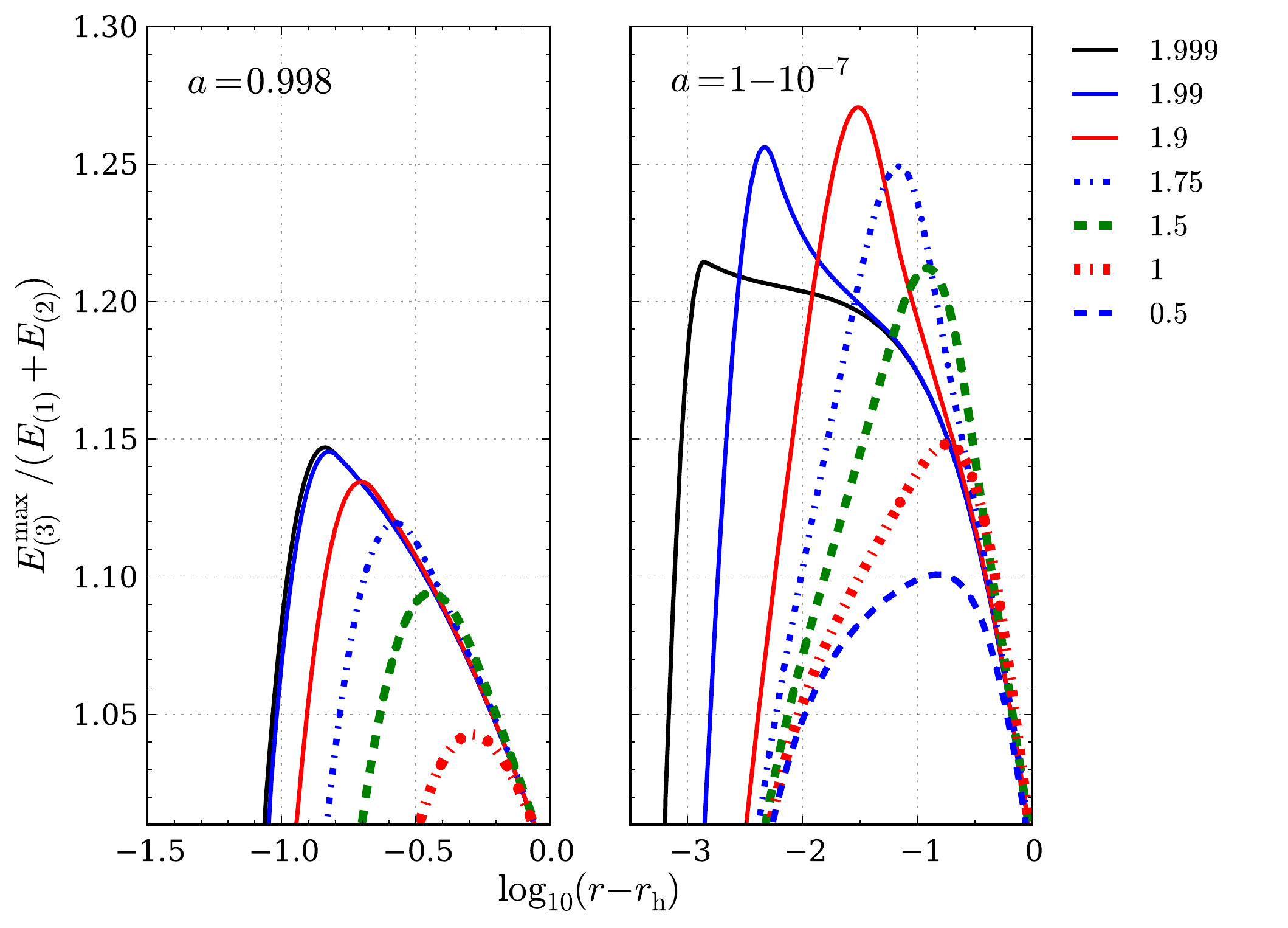}
\caption{The maximal energy for the escaping photon for $a<1$
($L_{(1)} = 2$, $L_{(2)} \in (0.5,1.999)$).
{\it Left panel}: Spin parameter $a=0.998$. {\it Right panel}: $a=1-10^{-7}$.}
\label{fig:adep}
\end{figure}

\begin{figure}[ht]
\includegraphics[width=0.5\textwidth]{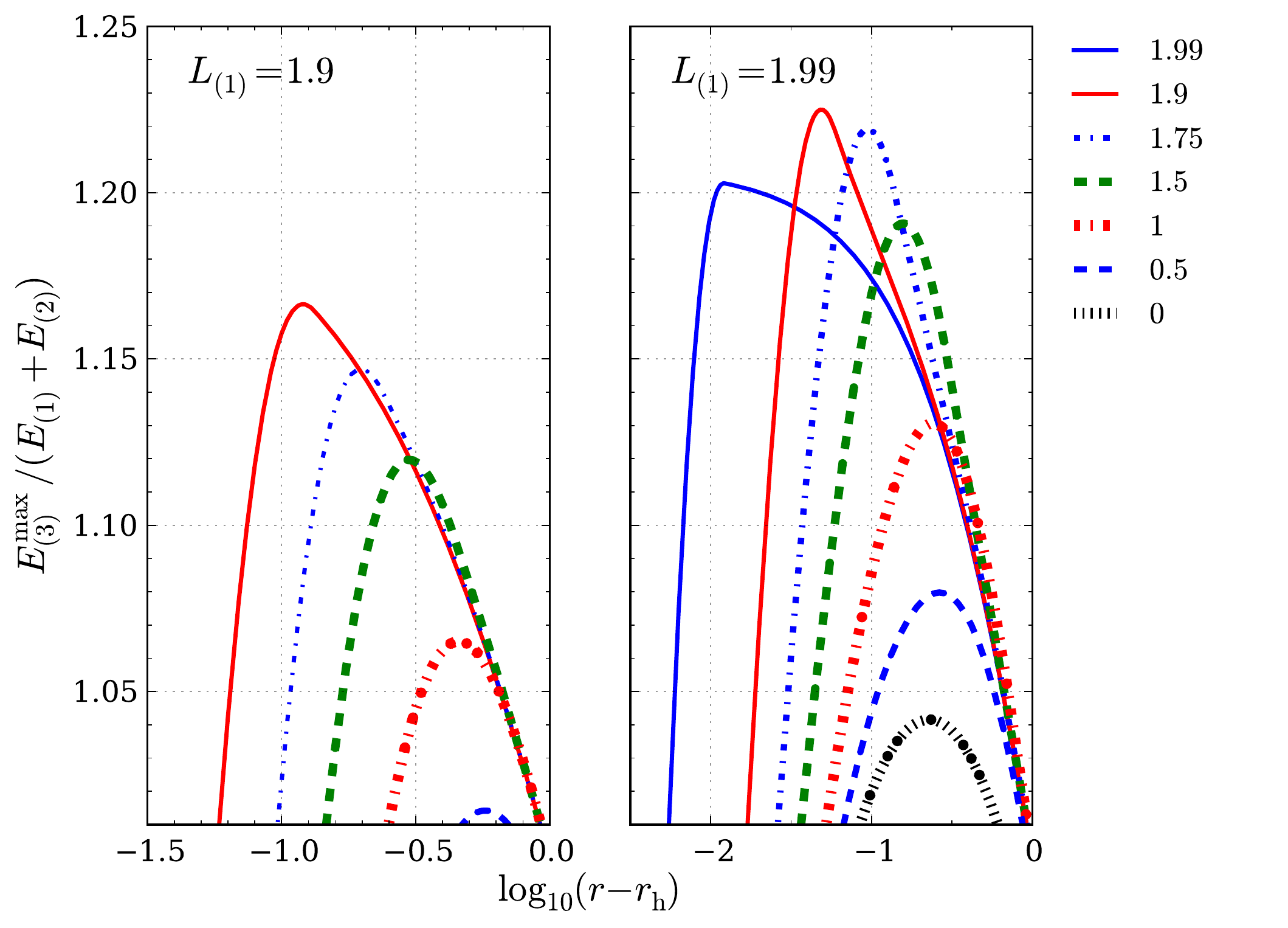}
\caption{Dependence of $E^{\rm max}_{(3)}$ on the maximal angular
momentum of one of the initially-infalling particles, $L_{(1)}$. Here,
$L_{(1)}=1.9$ (left panel) and $L_{(1)}=1.99$ (right panel), whereas
$L_{(2)}$ spans values not larger than $L_{(1)}$ ($a=1$).}
\label{fig:ldep}
\end{figure}

Fig.~\ref {fig:adep} depicts the dependence of $E^{\rm max}_{(3)}$ on the spin parameter
$a$. We keep $L_{(1)}=2$ and vary $L_{(2)}$ and the position of the collision; other
parameters are the same as for $a=1$. As expected, the maximal energies are smaller than in
the case of $a=1$. The self-similar behavior shown in Fig.~\ref {fig:a1l12} is recovered for
$L_{(2)}\to L_{(1)} = 2$ for $a\to 1$. Similarly, lowering the value of $L_{(1)}$ to values
smaller than the critical angular momentum results in lowering the $E^{\rm max}_{(3)}$
(Fig.~\ref{fig:ldep}).
%-----------------------------------------------------------------------------
%-----------------------------------------------------------------------------
%-----------------------------------------------------------------------------
%-----------------------------------------------------------------------------
%-----------------------------------------------------------------------------
%-----------------------------------------------------------------------------
%-----------------------------------------------------------------------------
%-----------------------------------------------------------------------------

%-----------------------------------------------------------------------------
%
%%%\section{Conclusions and implications}
%
%-----------------------------------------------------------------------------
We find that, for the specific case considered, a collision between two
massive particles falling from rest at infinity and colliding in the
equatorial plane, the maximal energy of an escaping photon is $\approx 1.3$
times the total initial energy at infinity of the infalling particles.
While there is an energy gain and energy is extracted from the BH, 
the energy gain is very modest, in stark contrast to the diverging CM
energy. Symmetry arguments suggest that for a configuration of two 
particles infalling from infinity, this is the maximal possible energy
gain even if the collision is not restricted to the equatorial plane.
Similarly one expects that the maximal energy gain will be lower if   the
escaping particle is massive.  We don't expect  the results to be
qualitatively different if we consider different initial conditions for
the infalling particles.

In retrospect it is easy to understand this result - large energy in the CM is not
sufficient; the energetic particles have to escape from the vicinity of the BH to
infinity.  This is demanding since the overall CM system has a diverging (comparable to the
large CM energy) negative radial momentum. For an outgoing photon to escape, it has to have
an outwards-pointing radial momentum (and such an angular momentum that it won't be deflected
back into the BH), or a sufficient angular momentum such that it will be deflected
before falling into the BH. Given the huge CM negative radial momentum, most of the 
resulting particles will move radially inwards and will be quickly swallowed by the BH. They 
won't have the angular momentum needed to turn around before reaching the horizon. 
Only in a very small regime of the parameter phase-space do the 
conditions allow the resulting photons to escape, but in this regime the
energy of the escaping photon is not large. Thus, the highly-energetic particles 
simply fall into the BH without any observable effects at infinity.

Are there any caveats that may enable us to avoid this conclusion and to have particles
with diverging energies escaping to infinity? As pointed out earlier, the main reason that
most resulting particles fall into the BH is the huge negative radial momentum of
the CM. In the optimal case, largest CM energies are attained when one of the particles
is just turning around near the horizon and the other one falls and collides with it.
Practically, the only way to avoid the negative radial momentum is when one of the two colliding  particles 
has a positive radial momentum. This is almost impossible as it would have to come
out of the BH. The orbit of an infalling particle that has just turned around would
be rather similar to the orbit of the turning particle and the overall CM energy won't be
large. One can speculate further and consider multiple collisions: the first collision
sends a particle outwards and in the second one it will collide with an infalling particle. 
Alas, the phase space for such a situation is extremely small and the process is simply unlikely.

It seems that from the point of view of a distant observer the CM energy is just an
`illusory' energy. One can imagine extremely energetic collisions very close to a BH, 
but the results of such collisions plunge quickly below the horizon, prohibiting 
a distant observer to know about them, let alone to detect energetic outgoing particles.

Our results confirm the earlier work of Piran and Shaham \cite{PS1977} that, while collisions
in the ergosphere can in principle enable us to extract energy from a rotating BH,
it is unlikely that the conditions for a significant energy extraction via a collisional
Penrose process would appear in nature. Similar considerations suggest that
hydrodynamical flows won't be efficient either. This leaves magnetic processes, such 
as the Blandford-Znajek mechanism, as the only viable way to extract rotational energy 
from a rotating BH. Indeed, there are now convincing arguments 
\cite{TchekhovskoyNM2011,NarayanM2012} indicating that the power needed to accelerate matter 
in the relativistic jets (with Lorentz factors up to $\gamma \approx 50$)  
observed in quasars and microquasars may be powered by the Blandford-Znajek mechanism, 
an electromagnetic version of the Penrose process.

%-----------------------------------------------------------------------------
{\it Acknowledgments.} This work was supported by the Polish 
National Science Centre grant no. UMO-2011/01/B/ST9/05437 and by an ERC 
Advanced Research grant (TP).  MB also acknowledges the support of the 
Polish MNiSW research grant no. N N203 512838.
%-----------------------------------------------------------------------------

\end{document}